\documentclass[english,reprint,amsmath,amssymb,aps,pra]{revtex4-1}
\usepackage[latin9]{inputenc}
\usepackage{amsmath}
\usepackage{graphicx}
\usepackage{esint}

\makeatletter

\providecommand{\tabularnewline}{\\}

\usepackage{dcolumn}
\usepackage{bm}

\usepackage{babel}

\usepackage{babel}

\usepackage{babel}

\usepackage{babel}

\makeatother

\usepackage{babel}
\begin{document}

\title{Spectrum, radial wave functions, and hyperfine splittings of the
Rydberg states in heavy-metal alkali atoms}

\author{Ali Sanayei and Nils Schopohl }

\email{nils.schopohl@uni-tuebingen.de}

\selectlanguage{english}%

\affiliation{Institut f{\"u}r Theoretische Physik and CQ Center for Collective
Quantum Phenomena and their Applications in LISA\textsuperscript{+},
Eberhard-Karls-Universit{\"a}t T{\"u}bingen, Auf der Morgenstelle
14, D-72076 T{\"u}bingen, Germany}

\affiliation{}

\date{\today}
\begin{abstract}
We present numerically accurate calculations of the bound state spectrum
of the highly excited valence electron in the heavy-metal alkali atoms
solving the radial Schr{\"o}dinger eigenvalue problem with a modern
spectral collocation method that applies also for a large principal
quantum number $n\gg1$. As an effective single-particle potential
we favor the reputable potential of Marinescu \emph{et al}., {[}Phys.
Rev. A \textbf{49}, 982 (1994){]}. Recent quasiclassical calculations
of the quantum defect of the valence electron agree for orbital angular
momentum $l=0,1,2,...$ overall remarkably well with the results of
the numerical calculations, but for the Rydberg states of rubidium
and also cesium with $l=3$ this agreement is less fair. The reason
for this anomaly is that in rubidium and cesium the potential acquires
for $l=3$ deep inside the ionic core a second classical region, thus
invalidating a standard WKB calculation with two widely spaced turning
points. Comparing then our numerical solutions of the radial Schr{\"o}dinger
eigenvalue problem with the uniform analytic WKB approximation of
Langer constructed around the remote turning point $r_{n,j,l}^{\left(+\right)}$
we observe everywhere a remarkable agreement, apart from a tiny region
around the inner turning point $r_{n,j,l}^{(-)}$. For \emph{s}-states
the centrifugal barrier is absent and no inner turning point exists,
$r_{n,j,0}^{(-)}=0$. With the help of an ansatz proposed by Fock
we obtain for the \emph{s}-states a second uniform analytic approximation
to the radial wave function complementary to the WKB approximation
of Langer, which is exact for $r\to0^{+}$. From the patching condition,
that for $l=0$ the Langer- and Fock solutions should agree in the
intermediate region $0<r\ll r_{n,j,l}^{\left(+\right)}$, not only
an equation determining the quasiclassical quantum defect $\delta_{0}$
but also the value of the radial \emph{s}-wave function at $r=0$
is analytically found, thus validating the Fermi-Segr{\`e} formula
for the hyperfine splitting constant $A_{n,j,0}^{\left(\mathrm{HFS}\right)}$.
As an application we consider recent spectroscopic data for the hyperfine
splittings of the isotopes $^{85}\mathrm{Rb}$ and $^{87}\mathrm{Rb}$
and find a remarkable agreement with the predicted scaling relation
$A_{n,j,0}^{\left(\mathrm{HFS}\right)}\left(n-\delta_{0}\right)^{3}=\mathrm{const}.$ 
\begin{description}
\item [{PACS numbers}] 31.10.+z, 31.15.-p, 31.30.Gs, 32.80.Ee {\small \par}

{\small \par}

\end{description}
\end{abstract}
\maketitle

\section{\label{Section I}introduction}

The alkali atoms have a simple ground state electronic structure,
with only one valence electron in an \emph{s}-state. On a level of
accuracy, where the relativistic corrections to the spectrum can be
ignored, the bound state spectrum of the excited valence electron
can be well described by the spherically symmetric effective single-particle
potential of Marinescu \emph{et al}. \cite{Marinescu et al,Units}:
\begin{equation}
V_{\mathrm{eff}}\left(r;l\right)=-2\frac{Z_{\mathrm{eff}}\left(r;l\right)}{r}-\alpha_{c}\frac{1-\exp\left(-\left(\frac{r}{r_{c}\left(l\right)}\right)^{6}\right)}{r^{4}}\,,\label{Marinescu et al.}
\end{equation}
where

\begin{equation}
Z_{\mathrm{eff}}\left(r;l\right)=1+\left(Z-1\right)e^{-ra_{1}\left(l\right)}-re^{-ra_{2}\left(l\right)}\left[a_{3}\left(l\right)+ra_{4}\left(l\right)\right]\,.\label{effective charge}
\end{equation}
This is actually a nonlocal potential, because it depends for each
proton number $Z$ of the alkali atom under consideration parametrically
on the orbital angular momentum $l=0,1,2,3,...$ of the valence electron.
At small distance $r$ to the atomic nucleus this effective interaction
potential mutates into a Coulomb potential, describing the interaction
of $Z$ protons with the outermost electron, and an additional (large)
constant; that is \cite{Units}, 
\begin{equation}
V_{\mathrm{eff}}\left(r;l\right)\rightarrow-\frac{2Z}{r}+2\left[\left(Z-1\right)a_{1}(l)+a_{3}(l)\right]\;\;\text{for\;\ }r\ll1\,.\label{effective potential very close to the origin-1}
\end{equation}
Conversely, far outside the ionic core region the potential converts
into a superposition of a long-ranged Coulomb term, describing the
interaction between a net positive charge $Z-\left(Z-1\right)=1$
and the valence electron (like in hydrogen atom), and a short-ranged
core polarization term; that is \cite{Units}, 
\begin{equation}
V_{\mathrm{eff}}\left(r;l\right)\rightarrow-\frac{2}{r}-\frac{\alpha_{c}}{r^{4}}\;\;\text{for\;\ }r\gg1\,.\label{effective potential far from the origin}
\end{equation}
In the region around the ionic core, comprising $Z-1$ strongly bound
electrons filling the inner electron shells of the atom, the two parameters
$\alpha_{c}$ and $r_{c}(l)$ represent the effects of the polarizability
of the latter, while the parameters $a_{1}(l)$, $a_{2}(l)$, $a_{3}(l)$,
and $a_{4}(l)$ shape the spatial dependence of the effective charge
$Z_{\mathrm{eff}}(r;l)$, as it alters as a function of $r$ from
unity to a value $Z$. For rubidium $Z=37$, for cesium $Z=55$, and
for francium $Z=87$.

Recently, a phenomenological modification of the potential for $l=1,2$
has been suggested in terms of a cutoff $r_{\mathrm{so}}\left(l\right)$
in the core region, which successfully predicts for all principal
quantum numbers $n$ and total angular momentum $j=l\pm1/2$ the fine
splittings of the Rydberg levels \cite{Units,Sanayei-Schopohl}:

\begin{equation}
V_{\mathrm{mod}}\left(r;j,l\right)=\begin{cases}
V_{\mathrm{eff}}\left(r;l\right) & \text{if }0\leqslant r\leqslant r_{\mathrm{so}}\left(l\right),\\
\\
V_{\mathrm{eff}}\left(r;l\right)+V_{\mathrm{SO}}\left(r;j,l\right) & \text{if }r>r_{\mathrm{so}}\left(l\right),
\end{cases}\label{modified potential}
\end{equation}
where $V_{\mathrm{SO}}\left(r;j,l\right)$ denotes the spin-orbit
potential. New precise spectroscopic data of $^{87}\mathrm{Rb}$ indeed
comply for all principal quantum numbers $n>7$ very well with the
(semi) analytical results obtained from quasiclassical WKB calculations,
cf. Tables I and II in Ref. \cite{Sanayei-Schopohl}.

In what follows, Sec. \ref{Section II}, we first check the accuracy
of our recent quasiclassical calculations of the spectrum of the highly
excited valence electron in $^{87}\mathrm{Rb}$ \cite{Sanayei-Schopohl}
with the potential (\ref{modified potential}), employing for the
solution of the radial Schr{\"o}dinger eigenvalue problem a modern
numerical collocation method based on the barycentric Chebyshev interpolation
\cite{Trefethen-Book,Trefethen_Matlab,Boyd_Book}. The results of
these numerical calculations indeed agree very well with our recent
quasiclassical calculations of the quantum defects for orbital angular
momentum $l=0,1,2$ and also $l\geq4$ , but for $l=3$ we spot for
the heavy-metal alkali atoms rubidium and cesium a discrepancy. In
Sec. \ref{Section III} we then provide an explanation for this discrepancy
bringing out for $l=3$ a hitherto unnoticed feature of the reputable
potential of Marinescu \emph{et al}. (\ref{Marinescu et al.}). In
Sec. \ref{Section IV} we show how to construct for the radial eigenfunctions
of the Rydberg states carrying an arbitrary orbital angular momentum
$l\geq0$ two complementary uniform quasiclassical approximations.
The first is the uniform WKB approximation of Langer \cite{Langer},
where we determine the normalization constant by the procedure described
by Bender and Orszag \cite{Bender-Orszag}. The obtained analytical
formula for the radial eigenfunctions of the Rydberg states for $l=0,1,2,...$
in fact agrees remarkably well with the numerical calculations almost
everywhere with exception of a small region around the origin at $r=0$.
Close to the origin, however, the Langer approximation becomes invalid.
We thus patch in the region well below the remote turning point the
quasiclassical approximation of Langer with an ansatz for the radial
wave function in terms of a Bessel function first proposed by Fock
\cite{Fock}, that is asymptotically exact for $r\rightarrow0^{+}$,
thus enabling us, for example for $l=0$ , to analytically determine
at the origin $r=0$ the value of the radial wave function for the
highly excited \emph{s}-states. In Sec. \ref{Section V}, finally,
we use these results to present a simple elementary proof for the
semi-empirical Fermi-Segr{\`e} formula \cite{Fermi and Segre} determining
the hyperfine splittings of the highly excited \emph{s}-states of
the alkali atoms.

\section{\label{Section II}spectral collocation on a chebyshev grid: a numerically
accurate method for the solution of the radial schr{\"o}dinger eigenvalue
problem}

To verify the accuracy of the quasiclassical calculations presented
in Ref. \cite{Sanayei-Schopohl} a numerically accurate method (see
supplementary material \cite{supplimentary}) is required, that solves
the radial Schr{\"o}dinger eigenvalue problem for the radial eigenfunctions
$R_{n,j,l}(r)=\frac{1}{r}\,U_{n,j,l}(r)$ with the modified potential
(\ref{modified potential}) reliably and accurately also for large
principal quantum numbers $n\gg1$ \cite{Units}:

\begin{equation}
\left[-\frac{\mathrm{d}^{2}}{\mathrm{d}r^{2}}+\frac{l\left(l+1\right)}{r^{2}}+V_{\mathrm{mod}}\left(r;j,l\right)-E_{n,j,l}\right]U_{n,j,l}(r)=0\,.\label{Schroedinger eigenvalue problem}
\end{equation}
To achieve this goal we use here a spectral collocation method \cite{Trefethen-Book,Trefethen_Matlab,Boyd_Book}
on a grid consisting of $k_{\mathrm{max}}+1$ Chebyshev grid points
obtained by projecting equally spaced points on the unit circle down
to the interval $\left[-1,1\right]$. Trivial scaling and shift leads
then to the not-equally spaced point set 
\begin{equation}
r_{k}=r_{\mathrm{max}}\,\frac{1-\cos\left(\pi\,\frac{k}{k_{\mathrm{max}}}\right)}{2},\;\;\;0\leqslant k\leqslant k_{\mathrm{ma}\mathrm{x}}\,,\label{Chebyshev grid}
\end{equation}
which clusters near $r=0$ and near $r=r_{\mathrm{max}}$. In sharp
contrast to a traditional finite difference method that controls the
error of numerical discretization by the choice of grid spacing, the
accuracy of a \emph{spectral} collocation method (a well-known concept
in modern numerical mathematics) is only limited by the smoothness
of the function being approximated \cite{Trefethen-Book,supplimentary}.
Implementing now spectral Chebyshev collocation the seeked wave function
$U_{n,j,l}(r)=rR_{n,j,l}\left(r\right)$ solving the radial Schr{\"o}dinger
eigenvalue problem (\ref{Schroedinger eigenvalue problem}) is represented
in terms of a finite vector $U_{n,j,l}\left(r_{k}\right)$ of its
values at the Chebyshev grid points $r_{k}$, thus defining implicitely
a stable and accurate Lagrange polynomial interpolant of degree $k_{\mathrm{max}}$.
Of particular value and simplicity is the numerically robust barycentric
representation of this interpolant due to Salzer \cite{Trefethen-Book}:

\begin{equation}
u_{n,j,l}(r)={\displaystyle \frac{{\textstyle \sum_{k=0}^{k_{\mathrm{max}}}}\frac{w_{k}\,U_{n,j,l}\left(r_{k}\right)}{r-r_{k}}}{\sum_{k'=0}^{k_{\mathrm{max}}}\frac{w_{k'}}{r-r_{k'}}}\,,}\label{Salzer formula}
\end{equation}
where

\begin{equation}
w_{k}=\left(-1\right)^{k}\times\begin{cases}
\frac{1}{2} & \text{if }k=0\;\textrm{or}\;k=k_{\mathrm{max}},\\
\\
1 & \text{otherwise. }
\end{cases}\label{weights}
\end{equation}
As a matter of fact, $u_{n,j,l}(r)$ is a polynomial of degree $k_{\mathrm{max}}$,
coinciding with the function values $U_{n,j,l}\left(r_{k}\right)$
at the grid points $r_{k}$. Well-known accuracy and stability concerns
regarding convergence of high order polynomial interpolants do not
apply to a Chebyshev grid with its not-equispaced points clustering
around the corner points of the grid \cite{Trefethen-Book}.

Replacing the function $U_{n,j,l}(r)$ by such a polynomial interpolant
$u_{n,j,l}(r)$ of degree $k_{\mathrm{max}}$ implies that derivative
operations on those functions are replaced by the same operations
applied to their interpolant. Thus, the first derivative $\frac{\mathrm{d}}{\mathrm{d}r}U_{n,j,l}(r)$
is now represented by a matrix $\mathbf{D}^{\left(1\right)}$ of size
$\left(k_{\mathrm{max}}+1\right)\times\left(k_{\mathrm{max}}+1\right)$
acting on the vector of function values $U_{n,j,l}\left(r_{k}\right)$
at $k_{\mathrm{max}}+1$ grid points $r_{k}$ \cite{Trefethen_Matlab},
likewise the second-order derivative $\frac{\mathrm{d}^{2}}{\mathrm{d}r^{2}}U_{n,j,l}(r)$
is represented by a matrix $\mathbf{D}^{\left(2\right)}=\mathbf{D}^{\left(1\right)}\circ\mathbf{D}^{\left(1\right)}$.
This approach converts the radial Schr{\"o}dinger eigenvalue problem
(\ref{Schroedinger eigenvalue problem}) into a standard matrix eigenvalue
problem.

A crucial point here is that in the calculations of the spectrum of
the highly excited bound valence electron the grid should be fine
enough to resolve the oscillations of the wave functions $U_{n,j,l}(r)$
under consideration also in the coarsest part of the grid in accordance
with the sampling theorem \cite{sampling theorem}. Moreover, the
largest grid point $r_{\mathrm{max}}$ should be located in the region
well beyond the remote classical turning point $r_{n,j,l}^{\left(+\right)}\simeq2/\left(-E_{n,j,l}\right)$,
say, $r_{\mathrm{max}}\simeq\frac{3}{2}r^{\left(+\right)}$. In effect,
one then requires Dirichlet boundary conditions for the eigenfunction
$U_{n,j,l}(r)$ at both ends of the grid: 
\begin{equation}
U_{n,j,l}(0)=0=U_{n,j,l}\left(r_{\mathrm{max}}\right)\,.\label{boundary conditions}
\end{equation}
These boundary conditions imply that the first and the last columns
as well as the first and the last row of the matrix $\mathbf{D}^{\left(2\right)}$
can be stripped off \cite{Trefethen_Matlab}, thus leading to a $\left(k_{\mathrm{max}}-1\right)\times\left(k_{\mathrm{max}}-1\right)$
matrix eigenvalue problem to be solved for the $k_{\mathrm{max}}-1$
unknown function values $U_{n,j,l}(r_{k})$ at the inner points of
the grid.

It should be noted that only eigenvectors with associated eigenvalue
$-1<E_{n,j,l}<0$ need to be searched \cite{Units}. Moreover, because
only eigenvectors with components $U_{n,j,l}(r_{k})$ becoming exponentially
small for $r_{k}$ well beyond the remote classical turning point
$r_{n,j,l}^{(+)}$ are meaningfull, all other solutions of the discrete
matrix eigenvalue problem being physically meaningless.

For a detailed discussion and demonstration of the accuracy of the
spectral collocation method on a Chebyshev grid, we refer to our supplementary
material \cite{supplimentary}, where we present a comparison with
the well-known analytical eigenfunctions of the hydrogen atom.

\begin{figure}[tp]
\begin{centering}
\includegraphics[scale=0.7]{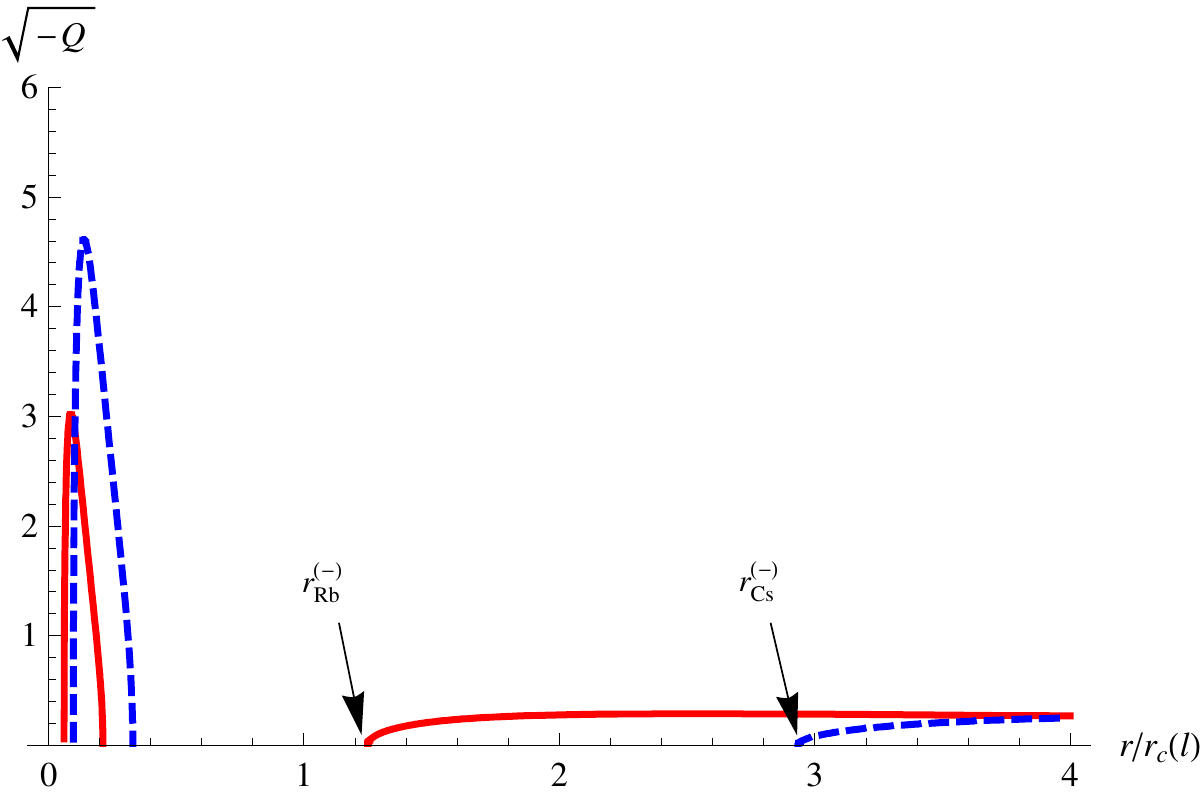} 
\par\end{centering}

\caption{(Color online) The quasiclassical momentum $\sqrt{-Q_{n,j,l}(r)}$
vs. scaled distance $r/r_{c}(l)$ for orbital angular momentum $l=3$
and total angular momentum $j=7/2$ of the excited bound valence electron
$\left(n\gg1\right)$ for rubidium (red) and cesium (dashed blue)
atoms, calculated with the effective potential of Marinescu \emph{et
al}. (\ref{Marinescu et al.}). There exists a tiny second classical
region located deep inside the atom core close to the origin, where
the quasiclassical momentum acquires again real values, well below
the positions of the inner turning points $r_{\mathrm{Rb}}^{(-)}$
and $r_{\mathrm{Cs}}^{(-)}$ for rubidium and cesium, respectively,
representing the lower boundary of their respective outer classical
regions extending up to their remote turning point $r_{n,j,l}^{\left(+\right)}$.
\protect \\
 }

\label{Fig 1. Quasiclassical momentum for L 3} 
\end{figure}

\begin{figure}[tp]
\begin{centering}
\includegraphics[scale=0.68]{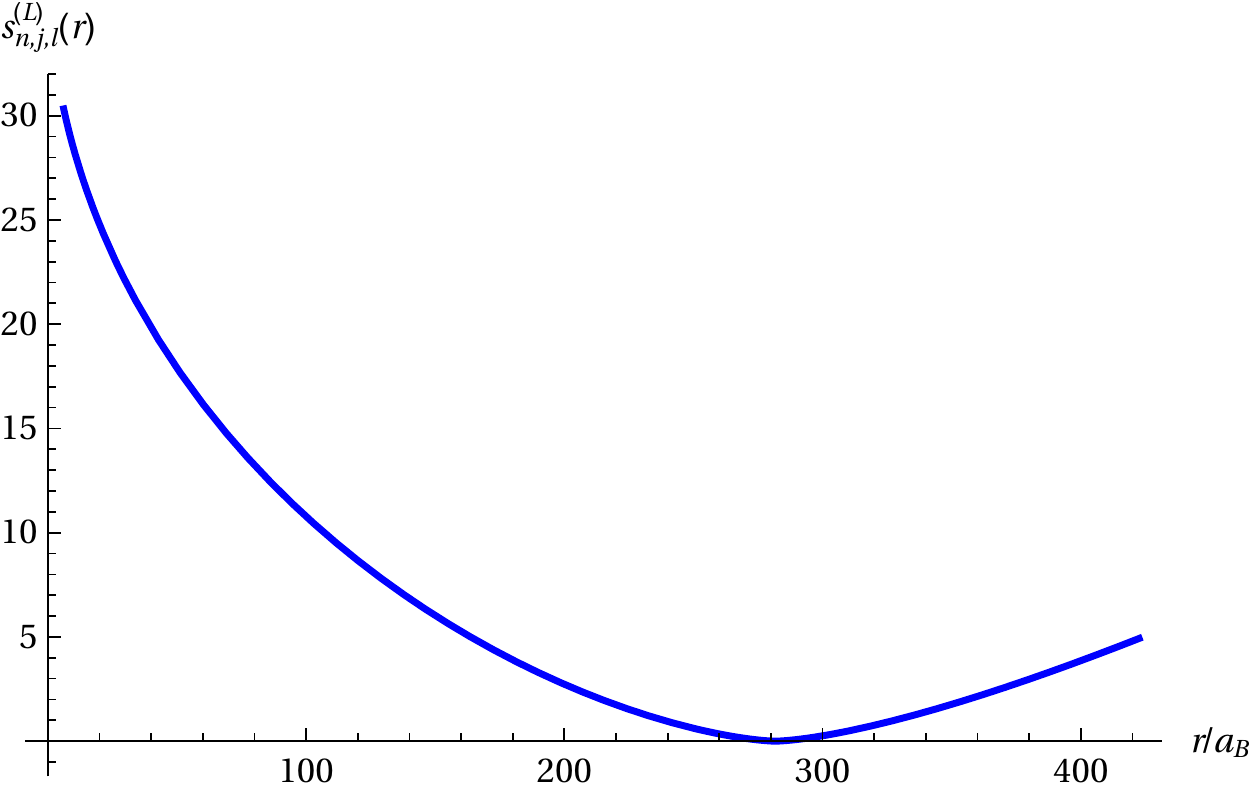} 
\par\end{centering}

\caption{(Color online) The Langer action integral $S_{n,j,l}^{(\mathrm{L})}(r)$,
cf. (\ref{Action integral}), as calculated from a barycentric polynomial
interpolant $s_{n,j,l}^{(\mathrm{L})}(r)$ on a Chebyshev grid, for
the excited bound valence electron of $^{87}\mathrm{Rb}$ with principal
quantum number $n=15$, orbital angular momentum $l=0$, and total
angular momentum $j=1/2$ . \protect \\
 }

\label{Fig 2. Langer action integral} 
\end{figure}

\section{\label{Section III}the quantum defect of the rydberg states in rubidium
and the $l=3$ anomaly in rubidium and cesium}

The bound state spectrum of the valence electron in $^{87}\mathrm{Rb}$,
as calculated by the aforementioned spectral collocation method \cite{supplimentary},
indeed agrees for almost all orbital angular momenta $l$, as well
with the spectroscopic data \cite{Li et al,Mack et al,Afroushed_g_states}
as with the quasiclassical calculations \cite{Sanayei-Schopohl},
with the exception of the $l=3$ Rydberg states \cite{Gallagher_f-states},
where a small systematic discrepancy is discernible between the results
obtained by the quasiclassical and the full numerical calculations,
cf. Table \ref{Table 1}. We offer here a simple explanation for this
anomaly, that applies only to the heavy alkali atoms rubidium and
cesium (and most likely also to francium), and which to the best of
our knowledge has not been reported before.

\begin{table*}
\qquad{}\caption{The values of quantum defect $\Delta_{j,l}$ associated with the Rydberg
level $n=15$ for $l=0,1,2,3,4$ and $j=l\pm1/2$. Experimental values
for $l=3,4$ are related to $^{85}\mathrm{Rb}$ and all theoretical
values correspond to $^{87}\mathrm{Rb}$. An estimation of uncertainties
for the values of quantum defect calculated by both quasiclaasical
theory and numerical collocation spectral method based on the barycentric
Chebyshev interpolation was obtained by varying the most effective
parameter of the reputable potential (\ref{Marinescu et al.}) to
the values of quantum defect {[}i.e., $a_{3}(l)${]} by around $1\%$.\protect \\
 }

\begin{tabular}{lcccccc}
\hline 
Quantum defect $\Delta_{j,l}$  & Expt. \cite{Li et al}  & Expt. \cite{Mack et al}  & Expt. \cite{Gallagher_f-states}  & Expt. \cite{Afroushed_g_states}  & Quasiclassical theory \cite{Sanayei-Schopohl}  & Numerical calculation (this work) \tabularnewline
\hline 
$\Delta_{1/2,0}$  & $3.132\,45(10)$  & $3.132\,45(2)$  & NA  & NA  & $3.131(3)$  & $3.132(3)$\tabularnewline
$\Delta_{1/2,1}$  & $2.656\,79(10)$  & NA  & NA  & NA  & $2.640(4)$  & $2.659(3)$\tabularnewline
\emph{$\Delta_{3/2,1}$}  & $2.643\,58(10)$  & NA  & NA  & NA  & $2.653(4)$  & $2.645(3)$\tabularnewline
$\left|\Delta_{1/2,1}-\Delta_{3/2,1}\right|$  & $0.013\,21(14)$  & NA  & NA  & NA  & $0.013(8)$  & $0.013(6)$\tabularnewline
$\Delta_{3/2,2}$  & $1.344\,86(4)$  & $1.344\,85(2)$  & NA  & NA  & $1.345(9)$  & $1.345(9)$\tabularnewline
$\Delta_{5/2,2}$  & $1.343\,27(3)$  & $1.343\,28(2)$  & NA  & NA  & $1.347(9)$  & $1.344(9)$\tabularnewline
$\left|\Delta_{3/2,2}-\Delta_{5/2,2}\right|$  & $0.001\,59(5)$  & $0.001\,57(3)$  & NA  & NA  & $0.001(18)$  & $0.001(18)$\tabularnewline
$\Delta_{5/2,3}$  & NA  & NA  & $0.016\,1406(9)$  & NA  & $0.013\,400(4)$  & $0.0164(4)$\tabularnewline
$\Delta_{7/2,3}$  & NA  & NA  & $0.016\,1606(7)$  & NA  & $0.013\,404(4)$  & $0.0164(4)$\tabularnewline
$\left|\Delta_{5/2,3}-\Delta_{7/2,3}\right|$  & NA  & NA  & $0.000\,0200(7)$  & NA  & $0.000\,004(8)$  & $0.000\,03(8)$\tabularnewline
$\Delta_{7/2,4}$  & NA  & NA  & NA  & $0.004\,05(6)$  & $0.005\,1500(4)$  & $0.003\,8385(4)$\tabularnewline
$\Delta_{9/2,4}$  & NA  & NA  & NA  & $0.004\,05(6)$  & $0.005\,1500(4)$  & $0.003\,8385(4)$\tabularnewline
\hline 
\end{tabular}

\label{Table 1} 
\end{table*}

There exists deep inside the atom core of rubidium and cesium, an
this applies as a matter of fact only for orbital angular momentum
$l=3$, a tiny second classical region of the potential (\ref{Marinescu et al.}),
see Fig. \ref{Fig 1. Quasiclassical momentum for L 3}, where the
classical (radial) momentum 
\begin{equation}
p_{n,j,l}\left(r\right)=\sqrt{-Q_{n,j,l}(r)}\,,\label{WKB momentum}
\end{equation}
with

\begin{equation}
Q_{n,j,l}(r)=\frac{l\left(l+1\right)}{r^{2}}+V_{\mathrm{mod}}\left(r;j,l\right)-E_{n,j,l}\,,\label{Q function}
\end{equation}
acquires as a function of distance $r$ to the origin again real-numbered
values. This feature invalidates a standard two-turning-point WKB
calculation of the spectrum of the $l=3$ Rydberg states, where the
widely spaced classical interval $r_{n,j,3}^{\left(-\right)}<r<r_{n,j,3}^{\left(+\right)}$
between the remote turning point $r_{n,j,3}^{\left(+\right)}$ and
the (second largest) inner turning point $r_{n,j,3}^{\left(-\right)}\ll r_{n,j,3}^{\left(+\right)}$
is taken into account, ignoring the existence of the tiny second classical
region inside the core of the atom for $l=3$, cf. Fig. \ref{Fig 1. Quasiclassical momentum for L 3}.
Because the asymptotics (\ref{effective potential very close to the origin-1})
of the potential reveals in the vicinity of the origin $r=0$ a large
constant term, which by far dominates the energy eigenvalues $E_{n,j,l}$
of the bound valence electron, the classical (radial) momentum inside
this second classical region is nearly independent on the energy variable
$-1<E_{n,j,l}<0$ of the bound states under consideration.

As explained in Ref. \cite{Sanayei-Schopohl}, the quantum defect
$\Delta_{j,l}=\delta_{l}+\eta_{j,l}$ is connected to the energy eigenvalue
$E_{n,j,l}$ of the bound valence electron with principal quantum
number $n\gg1$ and total angular momentum $j=l\pm1/2$ by \cite{Units,Gallagher_Book}
\begin{equation}
E_{n,j,l}=-\frac{1}{\left(n-\Delta_{j,l}\right)^{2}}\,,\label{energy eigenvalues}
\end{equation}
the fine splittings of the spectrum being thus to leading order proportional
to the difference $\Delta_{l-\frac{1}{2},l}-\Delta_{l+\frac{1}{2},l}=\eta_{l-\frac{1}{2},l}-\eta_{l+\frac{1}{2},l}$
of the associated quantum defects \cite{Sanayei-Schopohl}, cf. Table
\ref{Table 1}.

We find for all principal quantum numbers $n>7$ that choosing the
values of the cutoff $r_{\mathrm{so}}(l)$ in (\ref{modified potential})
according to the rule \cite{Units}

\begin{equation}
r_{\mathrm{so}}(l)\simeq\begin{cases}
0.0286294\times r_{c}\left(l\right)=0.043 & \text{for }l=1,\\
\\
0.0585394\times r_{c}\left(l\right)=0.285 & \text{for }l=2,\\
\\
0.135464\times r_{c}\left(l\right)=0.650 & \text{for }l=3,
\end{cases}\label{r_so values}
\end{equation}
the numerical calculations of the fine splitting agree surprisingly
well with the spectroscopic data of \cite{Li et al,Mack et al,Afroushed_g_states,Gallagher_f-states},
cf. Table \ref{Table 1}. Choosing larger or smaller values for $r_{\mathrm{so}}(l)$
than stated in (\ref{r_so values}), the calculated fine splittings
cease to give better agreement with experiment. Only for orbital angular
momentum $l=3$ we also find that changing the parameter $a_{3}(l)$
in the effective potential (\ref{Marinescu et al.}) from its tabulated
value in Ref. \cite{Marinescu et al} according to the rule $a_{3}(l=3)\rightarrow0.983431\times a_{3}\left(l=3\right)$
slightly improves the coincidence between the numerical calculations
and spectroscopic data \cite{Gallagher_f-states,Johnson et al}. Note
that for the quasiclassical calculation of quantum defect associated
with $l=0$ and $l=2$, we use the scaling prescription for $a_{3}\left(l=0\right)$
and $a_{3}\left(l=2\right)$ according to Ref. \cite{Sanayei-Schopohl}.

Recently, a calculation of the fine splittings for $l=3$ in rubidium
atoms has been carried out, taking a different potential and using
a relativistic many-body perturbation theory that employs relativistic
finite basis sets constructed from solutions to the single-electron
Dirac equation with a potential \cite{Blundell}. The results of these
calculations for the fine splittings of $l=3$ states in $\mathrm{Rb}$
atoms are closer to the experiments \cite{Gallagher_f-states,Brandenberger and Regal,Brandernberger and Malyshev}.
However, we should like to point out a serious consistency problem
attempting to solve a relativistic many-particle problem employing
a single-electron Dirac equation with a potential $V\left(r\right)$
that treats the relative coordinate $r$ as a four-vector, cf. Eq.
(1) in Ref. \cite{Blundell}. For a thourough analysis of the relativistic
H-atom, we refer to Ref. \cite{Barut}. A correct approach aiming
at taking into account the leading order of relativistic effects in
a many-electron problem should, in our opinion, be based on the Breit-Pauli
Hamiltonian \cite{Bethe and Salpeter,Froese_Fischer}, including not
only the usual spin-orbit term, but also the spin-spin interaction
term and the spin-other-orbit interaction \cite{Sanayei-Schopohl,Froese_Fischer}.
Both terms, the spin-spin interaction and the spin-other-orbit interaction,
influence the fine splitting as genuine relativistic multi-electron
terms which are certainly beyond the terms provided by any single-electron
Dirac equation, see Refs. \cite{Bethe and Salpeter,Froese_Fischer}
for expanded details. \\

\section{\label{Section IV}two complementary uniform quasiclassical approximations
for the radial eigenfunctions}

Once an energy eigenvalue $-1<E_{n,j,l}<0$ is determined from the
quasiclassical quantization condition, the corresponding uniform WKB
approximation of Langer to the solution of the radial Schr{\"o}dinger
equation (\ref{Schroedinger eigenvalue problem}), being constructed
around the remote turning point $r_{n,j,l}^{\left(+\right)}$, is
\cite{Langer,Bender-Orszag}:

\begin{widetext}

\begin{equation}
U_{n,j,l}^{\mathrm{(L)}}(r)=C_{n,j,l}^{(\mathrm{L})}\left[\frac{3}{2}S_{n,j,l}^{(\mathrm{L})}(r)\right]^{\frac{1}{6}}\left[\mathrm{sgn}\left(r-r_{n,j,l}^{\left(+\right)}\right)Q_{n,j,l}^{(\mathrm{L})}\left(r\right)\right]^{\frac{-1}{4}}\mathrm{Ai}\left(\mathrm{sgn}\left(r-r_{n,j,l}^{\left(+\right)}\right)\left[\frac{3}{2}S_{n,j,l}^{(\mathrm{L})}(r)\right]^{\frac{2}{3}}\right)\,.\label{Uniform Langer-WKB}
\end{equation}

\end{widetext}The function $\mathrm{Ai}(x)$ denotes the well-known
Airy function \cite{Bender-Orszag} and $\mathrm{sgn}(x)=\left|x\right|/x$.
The function $S_{n,j,l}^{(\mathrm{L})}(r)$ is the Langer action integral,

\begin{equation}
S_{n,j,l}^{(\mathrm{L})}(r)=\begin{cases}
\int_{r}^{r_{n,j,l}^{\left(+\right)}}\mathrm{d}r'\,\sqrt{-Q_{n,j,l}^{\left(\mathrm{L}\right)}(r')} & \text{if }r\leqslant r_{n,j,l}^{\left(+\right)},\\
\\
\int_{r_{n,j,l}^{\left(+\right)}}^{r}\mathrm{d}r'\,\sqrt{Q_{n,j,l}^{\left(\mathrm{L}\right)}(r')} & \text{if }r\geqslant r_{n,j,l}^{\left(+\right)},
\end{cases}\label{Action integral}
\end{equation}
where the function $\sqrt{-Q_{n,j,l}^{\left(\mathrm{L}\right)}(r)}$
is the quasiclassical momentum (\ref{WKB momentum}), but slightly
modified with the centrifugal barrier term being altered taking into
account the Langer correction $l\left(l+1\right)\rightarrow\left(l+\frac{1}{2}\right)^{2}$\cite{Langer_map}:

\begin{equation}
Q_{n,j,l}^{\left(\mathrm{L}\right)}(r)=\frac{\left(l+\frac{1}{2}\right)^{2}}{r^{2}}+V_{\mathrm{mod}}\left(r;j,l\right)-E_{n,j,l}\,.\label{Q Langer}
\end{equation}
For $l=0$ the centrifugal barrier term and the spin-orbit coupling
potential $V_{\mathrm{SO}}\left(r;j,l\right)$ are both absent, and
the lower turning point $r_{n,j,0}^{(-)}$ transforms into a singularity
of the radial Schr{\"o}dinger equation (\ref{Schroedinger eigenvalue problem}),
thus preventing a standard two-turning-point WKB calculation of the
spectrum. For a rigorous derivation of the normalization constant
$C_{n,j,l}^{(\mathrm{L})}$ we refer to Ref. \cite{Bender-Orszag}:

\begin{equation}
C_{n,j,l}^{(\mathrm{L})}=\left(-1\right)^{n-l-1}\,\sqrt{\frac{2\pi}{\int_{r_{n,j,l}^{\left(-\right)}}^{r_{n,j,l}^{\left(+\right)}}\frac{\mathrm{d}r}{\sqrt{-Q_{n,j,l}^{\left(\mathrm{L}\right)}(r)}}}}\,.\label{Normalization constant}
\end{equation}
In our WKB calculations we determine the positions $r=r_{n,j,l}^{\left(\pm\right)}$
of the turning points numerically by solving the implicit equation
$Q_{n,j,l}^{\left(\mathrm{L}\right)}(r)=0$. For large $n$ there
holds approximately

\begin{eqnarray}
r_{n,j,l}^{\left(+\right)} & \simeq & \begin{cases}
\frac{2}{-E_{n,j,l}} & \text{if }l=0,\\
\\
\frac{1}{-E_{n,j,l}}\left[1+\sqrt{1+\left(l+\frac{1}{2}\right)^{2}E_{n,j,l}}\,\right] & \text{if }l\geqslant1,
\end{cases}\label{outer turning point}
\end{eqnarray}
and

\begin{eqnarray}
r_{n,j,l}^{(-)} & \simeq & \begin{cases}
0 & \text{if }l=0,\\
\\
0.02\times r_{c}(l) & \text{if }l=1,2,\\
\\
\frac{\left(l+\frac{1}{2}\right)^{2}}{1+\sqrt{1+\left(l+\frac{1}{2}\right)^{2}E_{n,j,l}}} & \text{if }l\geqslant3.
\end{cases}\label{inner turning point}
\end{eqnarray}

In Fig. \ref{Fig 2. Langer action integral} the action integral $S_{n,j,l}^{(\mathrm{L})}(r)$
is displayed choosing, for example, $n=15$, $l=0$, and $j=1/2$.
Replacing the action integral (\ref{Action integral}) as a function
of the radial variable $r$ in (\ref{Uniform Langer-WKB}) by an accurate
barycentric interpolation polynomial on a suitable Chebyshev grid
(\ref{Chebyshev grid}), a substantial saving of computer time without
any loss of accuracy is attained. We found it advantageous to use
in the calculations of the action integral two complementary Chebyshev
grids, one with a number $k_{\mathrm{max}}$ of grid points $r_{k}$
in the interval $0\leqslant r_{k}\leqslant r_{n,j,l}^{\left(+\right)}$,
the other with a smaller number $k'_{\mathrm{ma}\mathrm{x}}$ of grid
points $r_{k'}$ in the interval $r_{n,j,l}^{\left(+\right)}\leqslant r_{k'}\leqslant r_{\mathrm{max}}$.

\begin{figure}
(a)

\begin{centering}
\includegraphics[scale=0.68]{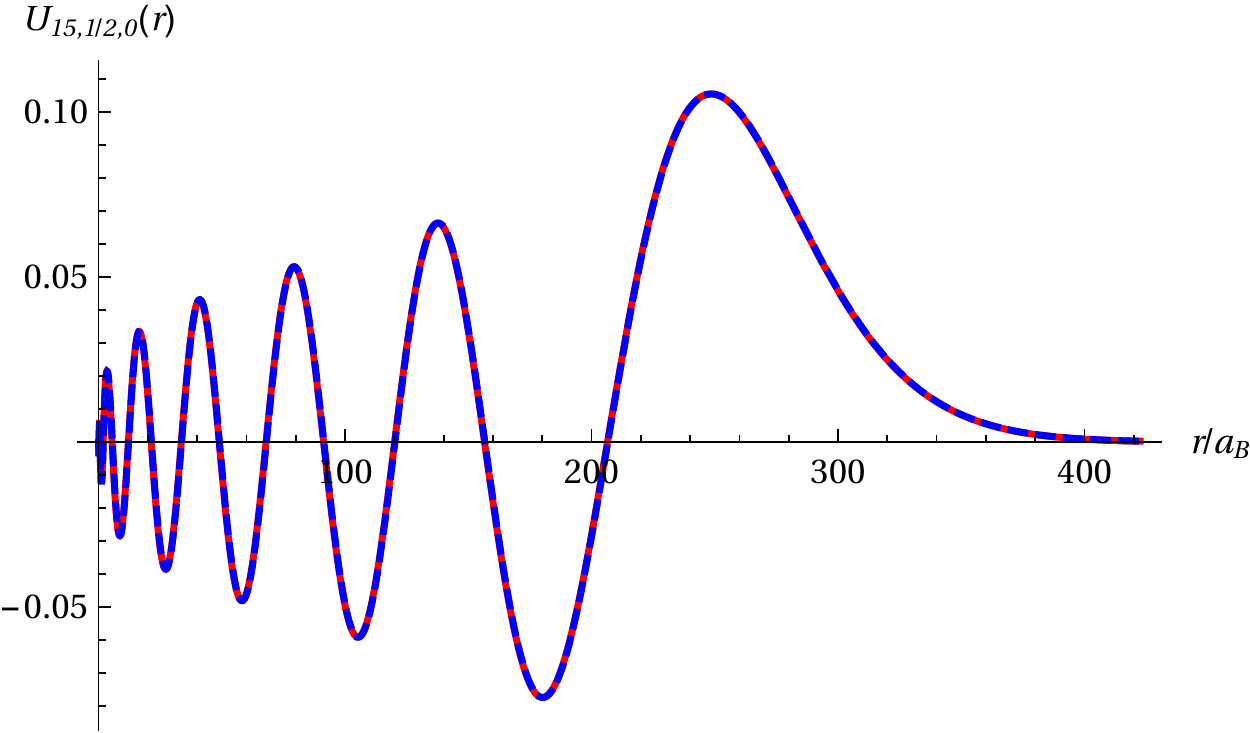} 
\par\end{centering}

(b)

\includegraphics[scale=0.68]{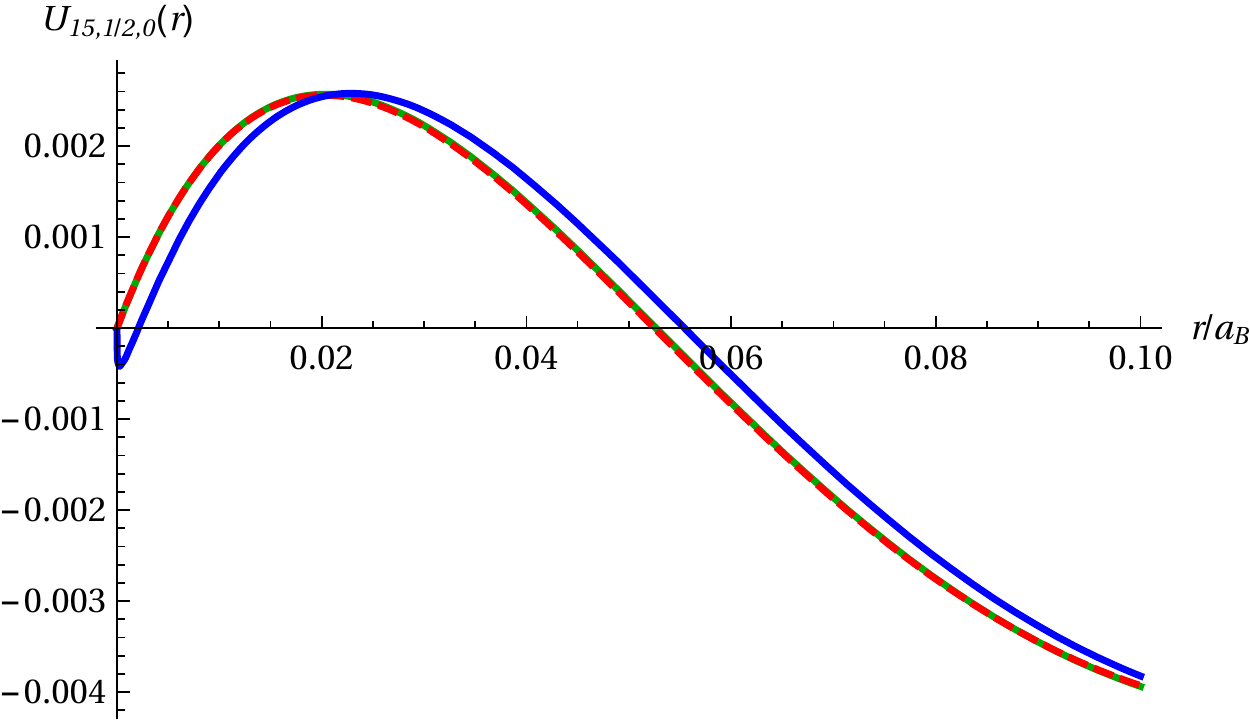}

\caption{(Color online) (a) The Chebyshev polynomial interpolant $u_{n,j,l}\left(r\right)$
to the eigenfunction $U_{n,j,l}\left(r\right)$ vs. radial distance
$r$ of the excited valence electron in $^{87}\mathrm{Rb}$ for principal
quantum number $n=15$, $l=0$, and $j=1/2$ as calculated with the
method of spectral collocation on a Chebyshev grid (red line), choosing
$r_{\mathrm{max}}=663.261$ and $k_{\mathrm{max}}=700$. Also shown
is the uniform WKB approximant $U_{n,j,l}^{\left(\mathrm{L}\right)}\left(r\right)$
of Langer (dashed blue), the error $\left|u_{n,j,l}\left(r\right)-U_{n,j,l}^{\left(\mathrm{L}\right)}\left(r\right)\right|$
being smaller than $10^{^{-3}}$ for $r>a_{B}$, cf. Ref. \cite{supplimentary};
(b) Expanded view around $r=0$ of $u_{n,j,l}\left(r\right)$ (dashed
red), of $U_{n,j,l}^{\left(\mathrm{L}\right)}\left(r\right)$ (blue)
and of the uniform quasiclassical approximant $U_{n,j,l}^{\left(\mathrm{F}\right)}\left(r\right)$
of Fock (green), the error $\left|u_{n,j,l}\left(r\right)-U_{n,j,l}^{\left(\mathrm{F}\right)}\left(r\right)\right|$
being smaller than $10^{^{-7}}$ for $r<3\times a_{B}$, for further
details see Ref. \cite{supplimentary}.\protect \\
 }

\label{Fig 3.Langer vs Numerics vs Fock} 
\end{figure}

In Fig. \ref{Fig 3.Langer vs Numerics vs Fock}(a), the (normalized)
Chebyshev polynomial interpolant $u_{n,j,l}\left(r\right)$ to the
radial eigenfunction $U_{n,j,l}\left(r\right)$ of the valence elelectron
of $^{87}\mathrm{Rb}$, as calculated from (\ref{Salzer formula})
with the method of spectral collocation on a Chebyshev grid, is plotted
for the excited valence electron in $^{87}\mathrm{Rb}$ for principal
quantum number $n=15$, $l=0$, and $j=1/2$ \cite{plot of higher excited states_footnote}.
With exception of a small region around the origin a remarkable agreement
is evident between the uniform WKB approximant $U_{n,j,l}^{\left(\mathrm{L}\right)}\left(r\right)$
of Langer and the Chebyshev polynomial approximant $u_{n,j,l}\left(r\right)$
to the eigenfunction $U_{n,j,l}\left(r\right)$.

Excluding a small region near to the lower boundary $r_{n,j,l}^{\left(-\right)}$
of the classically accessible region $r_{n,j,l}^{\left(-\right)}\leqslant r\leqslant r_{n,j,l}^{\left(+\right)}$,
the uniform WKB solution $U_{n,j,l}^{\left(\mathrm{L}\right)}\left(r\right)$
of Langer approximates for $r>r_{n,j,l}^{\left(-\right)}$ the exact
eigenfunction $U_{n,j,l}\left(r\right)$ of the valence electron in
the alkali atoms for arbitrary orbital angular momentum $l$ very
well, with the exception of the $l=3$ states in rubidium and cesium,
because of the second classically region inside the core, cf. Fig.
\ref{Fig 1. Quasiclassical momentum for L 3}. The key idea of the
uniform WKB approximation of Langer is to replace the spatial variation
of the potential around the turning points $r_{n,j,l}^{\left(\pm\right)}$
of the classically accessible region by a linear function of $r$,
thus reducing in that region the radial differential equation (\ref{Schroedinger eigenvalue problem})
to an analytically solvable one in terms of the Airy functions. But
$r_{n,j,l}^{\left(-\right)}$ is zero for $l=0$, and according to
(\ref{inner turning point}) it is very small for $l=1,2$. Hence,
for orbital angular momentum $l<4$ the spatial variation of the potential
(\ref{Marinescu et al.}), which is near to the origin a Coulomb potential,
cf. (\ref{effective potential very close to the origin-1}), in fact
cannot be approximated well by a linear function of $r$.

Fortunately, with the help of an ansatz proposed by Fock \cite{Fock},
a second uniform quasiclassical solution to (\ref{Schroedinger eigenvalue problem})
can be constructed, that approximates now close to the origin the
exact eigenfunction $U_{n,j,l}\left(r\right)$ very well, thus being
complementary to the uniform WKB solution (\ref{Uniform Langer-WKB}):

\begin{equation}
U_{n,j,l}^{(\mathrm{F})}(r)=\frac{C_{n,j,l}^{(\mathrm{F})}}{\sqrt{\frac{\mathrm{d}}{\mathrm{d}r}\ln\left[s_{n,j,l}(r)\right]}}\,J_{2l+1}\left(s_{n,j,l}(r)\right)\,.\label{Fock ansatz}
\end{equation}
Here $J_{k}(z)$ denotes a Bessel function of the order $k$, and
the unknown function $s_{n,j,l}(r)$ is chosen such that the differential
equation obeyed by the ansatz $U_{n,j,l}^{(\mathrm{F})}(r)$ in the
interval $0\leqslant r\ll r_{n,j,l}^{\left(+\right)}$ coincides with
the radial Schr{\"o}dinger equation (\ref{Schroedinger eigenvalue problem})
for $r\rightarrow0^{+}$, see Sec. \ref{Section V}.

Figure \ref{Fig 3.Langer vs Numerics vs Fock}(b) presents an expanded
view of the region around the origin, revealing that the uniform WKB
approximation of Langer ceases to agree well with the eigenfunction
$u_{n,j,l}\left(r\right)$ for $r\rightarrow0^{+}$. Instead, now
a remarkable agreement between $u_{n,j,l}\left(r\right)$ with the
uniform quasiclassical solution $U_{n,j,l}^{\left(\mathrm{F}\right)}\left(r\right)$,
as obtained with the ansatz of Fock, is evident.

\section{\label{Section V} quasiclassical wave functions and hyperfine splittings
of the rydberg \emph{s}-states }

\begin{figure}[tp]
\begin{centering}
\includegraphics[scale=0.68]{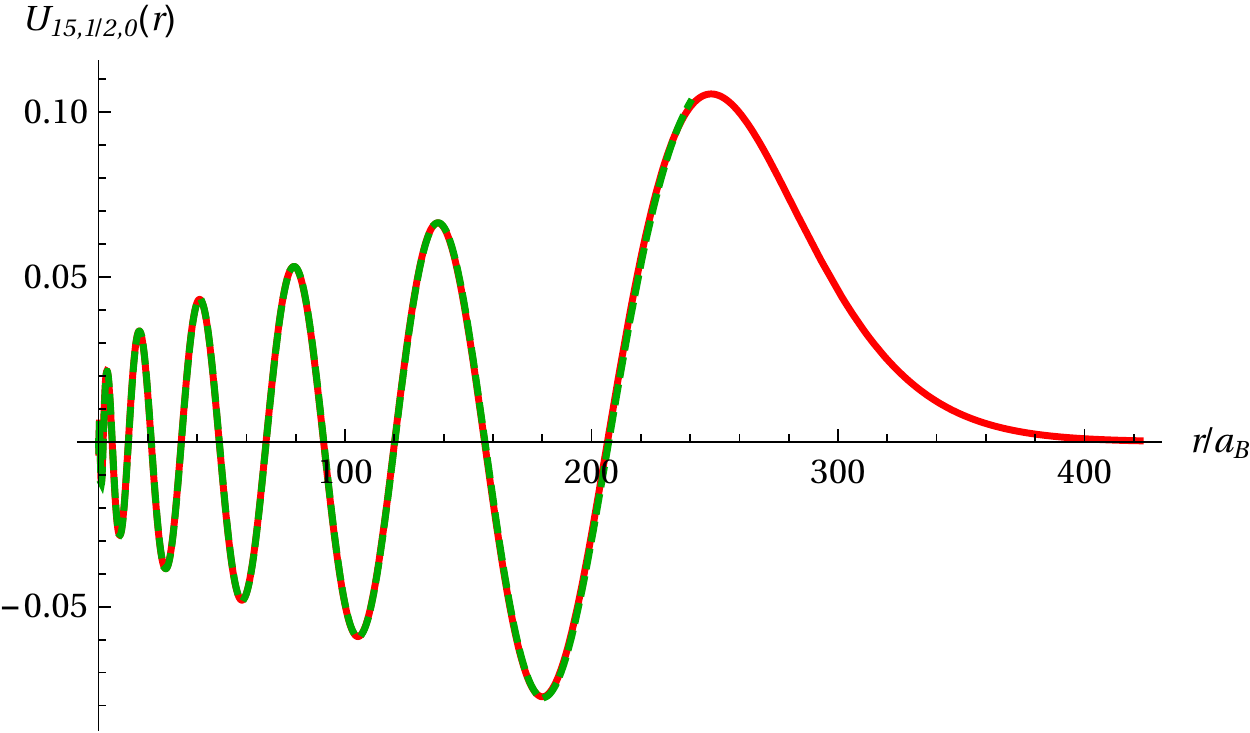} 
\par\end{centering}

\caption{(Color online) Comparison of the Chebyshev polynomial approximant
$u_{n,j,l}\left(r\right)$ to the normalized eigenfunction $U_{n,j,l}\left(r\right)$
as calculated with the method of spectral collocation on a Chebyshev
grid (red line), choosing $r_{\mathrm{max}}=663.261$ and $k_{\mathrm{max}}=700$,
with the uniform Fock ansatz (dashed green) associated with the bound
valence electron in $^{87}\mathrm{Rb}$ for the Rydberg level with
principal quantum number $n=15$, $l=0$, and $j=1/2$. \protect \\
 }

\label{Fig 4.Langer vs Fock} 
\end{figure}

We want to find out how the size of the hyperfine splittings of the
Rydberg \emph{s}-states depends on the principal quantum number $n$
and on the quantum defect $\delta_{0}$ . Due to the absence of the
centrifugal barrier and zero spin-orbit coupling for $l=0$ and $j=1/2$,
the associated exact radial wave function $U_{n,j,0}\left(r\right)=rR_{n,j,0}\left(r\right)$
solving the Schr{\"o}dinger eigenvalue problem (\ref{Schroedinger eigenvalue problem})
becomes near to the origin a linear function of $r$. Thus, it is
required that the quasiclassical aproximation $U_{n,j,0}^{(\mathrm{F})}(r)$
to $U_{n,j,0}\left(r\right)$ obeys to the boundary-value condition

\begin{equation}
\lim_{r\rightarrow0^{+}}\frac{U_{n,j,0}^{(\mathrm{F})}(r)}{r}=\lim_{r\rightarrow0^{+}}\frac{\mathrm{d}U_{n,j,0}^{(\mathrm{F})}(r)}{\mathrm{d}r}=R_{n,j,0}^{\left(\mathrm{F}\right)}\left(0\right)=\mathrm{const}.\label{Fock boundary condition}
\end{equation}
The task is to determine that constant $R_{n,j,0}^{\left(F\right)}\left(0\right)$
within the quasiclassical theory. A straightforward calculation shows
that the function $U_{n,j,0}^{(\mathrm{F})}(r)$ defined in (\ref{Fock ansatz})
solves the differential equation

\begin{equation}
\left[-\frac{\mathrm{d}^{2}}{\mathrm{d}r^{2}}+Q_{n,j,0}^{(\mathrm{F})}(r)\right]U_{n,j,0}^{(\mathrm{F})}(r)=0\,,\label{Fock differential eq}
\end{equation}
provided that

\begin{eqnarray}
Q_{n,j,0}^{(\mathrm{F})}(r) & = & -\left[s_{n,j,0}^{(1)}(r)\right]^{2}+\frac{3}{4}\left[\frac{s_{n,j,0}^{(1)}(r)}{s_{n,j,0}(r)}\right]^{2}\nonumber \\
 &  & +\frac{3}{4}\left[\frac{s_{n,j,0}^{(2)}(r)}{s_{n,j,0}^{(1)}(r)}\right]^{2}-\frac{1}{2}\,\frac{s_{n,j,0}^{(3)}(r)}{s_{n,j,0}^{(1)}(r)}\,.\label{Q Fock 1}
\end{eqnarray}
Here $f^{\left(k\right)}(r)\equiv\frac{\mathrm{d}^{k}}{\mathrm{d}r^{k}}f\left(r\right)$
denotes the derivative of order $k=1,2,3,...$ of a function $f\left(r\right)$.
The choice

\begin{equation}
s_{n,j,0}(r)=S_{n,j,0}^{(\mathrm{F})}\left(r\right)\equiv\int_{0}^{r}\mathrm{d}r'\,\sqrt{-Q_{n,j,0}(r')}\,,\label{S Fock}
\end{equation}
with 
\begin{equation}
Q_{n,j,0}(r)=V_{\mathrm{eff}}\left(r;l=0\right)-E_{n,j,0}\,,
\end{equation}
leads now to the identification

\begin{eqnarray}
Q_{n,j,0}^{(\mathrm{F})}(r) & = & Q_{n,j,0}(r)-\frac{3}{4}\,\frac{Q_{n,j,0}(r)}{\left[S_{n,j,0}^{(\mathrm{F})}\left(r\right)\right]^{2}}\nonumber \\
 &  & +\frac{5}{16}\left[\frac{Q_{n,j,0}^{(1)}\left(r\right)}{Q_{n,j,0}\left(r\right)}\right]^{2}-\frac{1}{4}\,\frac{Q_{n,j,0}^{(2)}\left(r\right)}{Q_{n,j,0}\left(r\right)}\,.\label{Q Fock 2}
\end{eqnarray}
For $r\rightarrow0^{+}$ the residue vanishes, that is $\frac{Q_{n,j,0}^{(\mathrm{F})}(r)-Q_{n,j,0}(r)}{Q_{n,j,0}(r)}$$\rightarrow0$,
implying that the Fock ansatz (\ref{Fock ansatz}) represents for
$l=0$ inside the classically accessible interval $0\leqslant r<r_{n,j,0}^{\left(+\right)}$
a second uniform approximation to the solution of the radial Schr{\"o}dinger
equation (\ref{Schroedinger eigenvalue problem}). The uniform quasiclassical
solution of Fock, which we present for $l=0$ now in the guise

\begin{equation}
U_{n,j,0}^{(\mathrm{F})}(r)=C_{n,j,0}^{(\mathrm{F})}\,\frac{\sqrt{S_{n,j,0}^{\left(\mathrm{F}\right)}\left(r\right)}}{\left[-Q_{n,j,0}\left(r\right)\right]^{\frac{1}{4}}}\,J_{1}\left(S_{n,j,0}^{\left(\mathrm{F}\right)}\left(r\right)\right)\,,\label{U Fock}
\end{equation}
indeed approximates inside the classically accessible region $0\leqslant r<r_{n,j,0}^{\left(+\right)}$
the exact eigenfunctions $U_{n,j,0}\left(r\right)$ of the Rydberg
\emph{s}-states of the valence electron in the alkali atoms very well,
almost up to the remote turning point $r_{n,j,0}^{\left(+\right)}$,
see Fig. \ref{Fig 4.Langer vs Fock}.

Deep inside the classically accessible region $0\ll r\ll r_{n,j,0}^{(+)}$
both action integrals, $S_{n,j,0}^{(\mathrm{F})}\left(r\right)$ and
$S_{n,j,0}^{(\mathrm{L})}(r)$, see (\ref{S Fock}) and (\ref{Action integral}),
assume for $n\gg1$ large values, so that the well-known asymptotics
of the Bessel function $J_{1}\left(z\right)$ and of the Airy function
$\mathrm{Ai}\left(-z\right)$ , valid for large arguments $z\gg1$
, can be used \cite{NIST}: 
\begin{equation}
\mathrm{Ai}\left(-z\right)\rightarrow\frac{1}{\sqrt{\pi}}\,\frac{\cos\left(\frac{2}{3}z^{\frac{3}{2}}-\frac{\pi}{4}\right)}{z^{\frac{1}{4}}}\,,\label{Airy asymp}
\end{equation}
and 
\begin{equation}
J_{1}\left(z\right)\rightarrow\sqrt{\frac{2}{\pi z}}\,\cos\left(z-\frac{3}{4}\pi\right)\,.\label{Bessel asymp}
\end{equation}
Accordingly, the uniform approximations of Langer (\ref{Uniform Langer-WKB})
and of Fock (\ref{U Fock}) respectively simplify in that region to

\begin{equation}
U_{n,j,0}^{\mathrm{(L)}}(r)\rightarrow\frac{C_{n,j,0}^{(\mathrm{L})}}{\sqrt{\pi}}\,\frac{\cos\left(S_{n,j,0}^{\left(\mathrm{L}\right)}\left(r\right)-\frac{\pi}{4}\right)}{\left[-Q_{n,j,0}\left(r\right)\right]^{\frac{1}{4}}}\,,
\end{equation}
and 
\begin{equation}
U_{n,j,0}^{\mathrm{(F)}}(r)\rightarrow C_{n,j,0}^{(\mathrm{F})}\,\sqrt{\frac{2}{\pi}}\,\frac{\cos\left(S_{n,j,0}^{\left(\mathrm{F}\right)}\left(r\right)-\frac{3}{4}\pi\right)}{\left[-Q_{n,j,0}\left(r\right)\right]^{\frac{1}{4}}}\,.
\end{equation}
The patching requirement that both functions $U_{n,j,0}^{\mathrm{(L)}}(r)$
and $U_{n,j,0}^{\mathrm{(F)}}(r)$ should coincide for $0\ll r\ll r_{n,j,0}{}^{\left(+\right)}$
can only be fulfilled provided that

\begin{equation}
S_{n,j,0}^{\left(\mathrm{F}\right)}\left(r\right)+S_{n,j,0}^{\left(\mathrm{L}\right)}\left(r\right)=\int_{0}^{r_{n,j,0}^{(+)}}\mathrm{d}r'\,\sqrt{-Q_{n,j,0}\left(r'\right)}=n\pi\,,\label{eq:quasiclassical quantization l=00003D00003D00003D0}
\end{equation}
and 
\begin{equation}
C_{n,j,0}^{(\mathrm{F})}=\frac{\left(-1\right)^{n-1}}{\sqrt{2}}\,C_{n,j,0}^{(\mathrm{L})}\,.\label{eq:normalization constant Fock}
\end{equation}
Equation (\ref{eq:quasiclassical quantization l=00003D00003D00003D0})
is the quasiclassical quantization condition for zero orbital angular
momentum $l=0$ \cite{Sanayei-Schopohl,Migdal}, determining here
the energy levels of the Rydberg \emph{s}-states \cite{Units} 
\begin{equation}
E_{n,j,0}=-\frac{1}{\left(n-\delta_{0}\right)^{2}}\,,\label{eq:quantum defect l=00003D00003D00003D0}
\end{equation}
with $\delta_{0}\equiv\Delta_{1/2,0}$ the quantum defect of the valence
electron for $l=0$. It turns out that (\ref{U Fock}) is a very good
approximation to the eigenfunction $U_{n,j,0}(r)$ everywhere in the
classically accessible region below the remote turning point, cf.
Fig. \ref{Fig 4.Langer vs Fock}. 

The normalization constant (\ref{Normalization constant}) for $l=0$
can also be expressed analytically in terms of the quantum defect
$\delta_{0}$. To see this, let us write for the moment being the
remote turning point $r_{n,j,0}^{(+)}$ as a function of the energy
variable $E$, i.e., the function $r^{\left(+\right)}\left(E\right)$
is determined from the requirement 
\begin{equation}
E-V_{\mathrm{eff}}\left(r^{\left(+\right)}(E);l=0\right)=0\,.
\end{equation}
We now rewrite (\ref{Normalization constant}) in the guise 
\begin{equation}
\frac{1}{\left|C_{n,j,0}^{\left(\mathrm{L}\right)}\right|^{2}}=\lim_{E\rightarrow E_{n,j,0}}\frac{\mathrm{d}}{\mathrm{d}E}\nu\left(E\right)\,,
\end{equation}
with $\nu\left(E\right)$ denoting the action integral \cite{Sanayei-Schopohl}
\begin{equation}
\nu\left(E\right)=\frac{1}{\pi}\int_{0}^{r^{\left(+\right)}\left(E\right)}\mathrm{d}r'\,\sqrt{E-V_{\mathrm{eff}}\left(r';l=0\right)}\,.
\end{equation}
With the help of the relation $\frac{\mathrm{d}}{\mathrm{d}E}\nu\left(E\right)=\frac{1}{\mathrm{d}E/\mathrm{d}\nu}$
and taking into account the identity $\lim_{E\rightarrow E_{n,j,0}}\nu\left(E\right)=n$,
cf. (\ref{eq:quasiclassical quantization l=00003D00003D00003D0}),
there follows from (\ref{eq:normalization constant Fock}) at once
for $l=0$ and $j=1/2$: 
\begin{equation}
\left|C_{n,j,0}^{\left(\mathrm{F}\right)}\right|^{2}=\frac{1}{2}\left|C_{n,j,0}^{\left(\mathrm{L}\right)}\right|^{2}=\frac{1}{2}\,\frac{\mathrm{d}}{\mathrm{d}n}E_{n,j,0}=\frac{1-\frac{\mathrm{d}}{\mathrm{d}n}\delta_{0}}{\left(n-\delta_{0}\right)^{3}}\,.\label{eq: normalization constant Fock II}
\end{equation}

\begin{table*}[tp]
\qquad{}\caption{Values of the scaled magnetic dipole interaction (hyperfine splitting)
constant $\frac{A_{n,j,0}^{\left(\mathrm{HFS}\right)}}{h}\left(n-\delta_{0}\right)^{3}$,
in gigahertz, associated with the highly excited \emph{s}-states of
the bound valence electron in $^{85}\mathrm{Rb}$ and $^{87}\mathrm{Rb}$.
Experiments \cite{Li et al} and \cite{Mack et al} were carried out
for principal quantum numbers $n\in\left\{ 28,29,30,31,32,33\right\} $,
and Experiment \cite{Tauschinsky et al} for $n\in\left\{ 20,21,22,23,24\right\} $.
Note that an estimation of the uncertainty for the numerical calculation
of quantum defect $\delta_{0}$ (cf. Table \ref{Table 1}) was obtained
by varying the most effective parameter of the reputable potential
(\ref{Marinescu et al.}) to the values of quantum defect {[}i.e.,
$a_{3}(0)${]} by around $1\%$.\protect \\
 }

\begin{tabular}{ccccc}
\hline 
$\;\;\;\;\;$Isotope$\;\;\;\;\;$  & $\;\;\;\;\;$Expt. \cite{Li et al}$\;\;\;\;\;$  & $\;\;\;\;\;$Expt. \cite{Mack et al}$\;\;\;\;\;$  & $\;\;\;\;\;$Expt. \cite{Tauschinsky et al}$\;\;\;\;\;$  & $\;\;\;\;\;$Theory (this work)$\;\;\;\;\;$ \tabularnewline
\hline 
$^{85}\mathrm{Rb}$  & $4.87(14)$  & NA  & NA  & $5.082(3)$\tabularnewline
$^{87}\mathrm{Rb}$  & NA  & $16.75(9)$  & $18.55(2)$  & $17.223(3)$\tabularnewline
\hline 
\end{tabular}

\label{Table 3} 
\end{table*}

For the $l=0$, $j=1/2$ states of the valence electron in the alkali
atoms the fine structure splitting due to spin-orbit coupling (assuming
exact spherical symmetry of the effective potential) is zero. Neglecting
the electric quadrupole moment of the nucleus a detectable shift in
the spectrum can now be attributed to the hyperfine interaction of
the magnetic moment of the valence electron with the nuclear magnetic
moment \cite{Arimondo}. Within the range of validity of the Fermi-contact-interaction
model, the size of the spectral splitting is then determined by the
magnetic dipole interaction (hyperfine splitting) constant \cite{Arimondo}

\begin{eqnarray}
A_{n,j,0}^{\left(\mathrm{HFS}\right)} & = & \frac{2}{3}\,\mu_{0}g_{s}\widetilde{g}_{I}\mu_{B}^{2}\,\lim_{\left|\mathbf{r}\right|\rightarrow0^{+}}\left|\psi_{n,j,0}\left(\mathbf{r}\right)\right|^{2}\,.\label{hyperfine constant}
\end{eqnarray}
Here $\mu_{0}$ is the vacuum permeability, $\mu_{B}=\frac{\left|e\right|\hbar}{2m_{e}}$
denotes the Bohr magneton, and the \emph{g}-factors of electron and
nucleus are $g_{s}=2.0023193043622$ and $\widetilde{g}_{I}=\frac{m_{e}}{m_{p}}g_{I}$,
respectively. For $^{87}\mathrm{Rb}$ it is found that $\widetilde{g}_{I}=-0.0009951414$,
and for $^{85}\mathrm{Rb}$, $\widetilde{g}_{I}=-0.00029364000$ \cite{Rb data sheet}.
It should be noted that in our system of units, see \cite{Units},
the particle density distribution $\left|\psi_{n,j,0}\left(\mathbf{r}\right)\right|^{2}$
is being measured as the number of particles per unit volume $\left(a_{B}\right)^{3}$.

The value of the wave function of the Rydberg \emph{s}-states $\psi_{n,j,0}\left(\mathbf{r}\right)=R_{n,j,0}(r)\,Y_{0,0}\left(\vartheta,\varphi\right)$
at the origin $r=0$ can be calculated analytically using the asymptotics
of the action integral (\ref{S Fock}) for small $r$: 
\begin{equation}
S_{n,j,0}^{\left(\mathrm{F}\right)}\left(r\right)\rightarrow\sqrt{8Zr}+\mathcal{O}\left(r^{\frac{3}{2}}\right)\,.\label{eq:Fock action for small r}
\end{equation}
Insertion of (\ref{eq:Fock action for small r}) into (\ref{U Fock})
leads then together with the analytical result (\ref{eq: normalization constant Fock II})
for the normalization constant to the exact result

\begin{eqnarray}
\lim_{\left|\mathbf{r}\right|\rightarrow0^{+}}\left|\psi_{n,j,0}\left(\mathbf{r}\right)\right|^{2} & = & \lim_{r\rightarrow0^{+}}\left|\frac{U_{n,j,0}^{(\mathrm{F})}\left(r\right)}{r}\frac{1}{\sqrt{4\pi}}\right|^{2}\nonumber \\
 & = & \frac{Z}{\pi}\frac{1-\frac{\mathrm{d}}{\mathrm{d}n}\delta_{0}}{\left(n-\delta_{0}\right)^{3}}\,.\label{value of psi at theorigin}
\end{eqnarray}
This formula connects the value of the \emph{$s$}-state wave function
at the origin to the derivative $\frac{\mathrm{d}}{\mathrm{d}n}E_{n,j,0}$
of the bound state spectrum in a radial Schr{\"o}dinger eigenvalue
problem. In the literature it is is often referred to as the semi-empirical
Fermi-Segr{\`e} formula \cite{Fock,Fermi and Segre,Prestage et al}.
For a rigorous derivation for differential equations of the type (\ref{Schroedinger eigenvalue problem}),
based on an identity for the Wronski determinant, see Ref. \cite{Durand}.

Equation (\ref{hyperfine constant}) engenders that the magnetic dipole
interaction (hyperfine splitting) constant $A_{n,j,0}^{\left(\mathrm{HFS}\right)}$
for the highly excited valence electron of the alkali atoms $\left(\mathrm{d}\delta_{0}/\mathrm{d}n\approx0\right)$
indeed should obey to the scaling relation 
\begin{equation}
A_{n,j,0}^{\left(\mathrm{HFS}\right)}\left(n-\delta_{0}\right)^{3}=\mathrm{const}.\label{eq:hyperfine scaling relation}
\end{equation}
In experiment the hyperfine level shift depends on nuclear spin $I$,
total angular momentum of the valence electron $j$, and on total
angular momentum $F$ assuming values in the interval $|I-j|\leqslant F\leqslant I+j$.
If only the magnetic dipole interaction was considered, then for $l=0$
, $j=1/2$ a level $E_{n,j,0}$ would split as a result of the magnetic
hyperfine interaction for the special case of nuclear spin $I\geqslant1/2$
into a doublet structure with quantum numbers $F=I\pm1/2$ \cite{Arimondo}:
\begin{equation}
\Delta E_{n,j,0}^{\left(\mathrm{HFS}\right)}=A_{n,j,0}^{\left(\mathrm{HFS}\right)}\,\frac{F\left(F+1\right)-I\left(I+1\right)-j\left(j+1\right)}{2}\,.
\end{equation}

Table \ref{Table 3} compares the theoretical values of the magnetic
dipole interaction (hyperfine splitting) constant $A_{n,j,0}^{\left(\mathrm{HFS}\right)}$
obtained from (\ref{eq:hyperfine scaling relation}) for $^{85}\mathrm{Rb}$
and $^{87}\mathrm{Rb}$ atoms with spectroscopic data \cite{Li et al,Mack et al,Tauschinsky et al}.
Overall, a very good agreement between theory and experiment can be
observed.

\section{\label{Conclusions}conclusions}

Using a numerically accurate and easy to implement modern numerical
method, namely, spectral collocation on a Chebyshev grid \cite{Trefethen-Book,Trefethen_Matlab,Boyd_Book}
based on the barycentric interpolation formula of Salzer (\ref{Salzer formula}),
we solved the radial Schr{\"o}dinger eigenvalue problem and determined
the excitation spectrum of the bound valence electron in the alkali
atoms, thus confirming the high accuracy of recent quasiclasscial
calculations of the quantum defect for the Rydberg states carrying
orbital angular momentum $l=0,1,2$ or $l>3$, with exception of the
$l=3$ Rydberg states of rubidium and cesium atoms. As a reason for
this anomaly we identified as a feature of the potential of Marinescu
\emph{et al}. \cite{Marinescu et al}, existing only for orbital angular
momentum $l=3$, a tiny second classical region located deep inside
the atom core around the nucleus of alkali atoms with proton number
$Z\geqslant37$, cf. Fig. (\ref{Fig 1. Quasiclassical momentum for L 3}),
thus invalidating for the heavy alkali atoms, rubidium and cesium
(and possibly also francium), a standard WKB calculation with only
two widely spaced turning points. Also, we found that the uniform
WKB approximation of Langer for the radial wave function of the valence
electron for $l\neq3$ indeed represents almost everywhere a remarkably
accurate approximation to the exact solution of the radial Schr{\"o}dinger
eigenvalue problem (\ref{Schroedinger eigenvalue problem}), omitting
a tiny interval near to the lower turning point of the classically
accessible region. In the region around the origin, where the unifrom
WKB approximation of Langer for the \emph{$s$}-states ceases to be
valid, we then showed using an ansatz of Fock \cite{Fock}, that a
complementary uniform quasiclassical solution in terms of a Bessel
function can be constructed, that coincides with the exact solution
of the radial wavefunction for $r\rightarrow0^{+}$. The uniform quasiclassical
approximation of Fock for the Rydberg the \emph{s}-states was found
to approximate the exact radial eigenfunction almost everywhere in
the classically accessible region remarkably well, with exception
of a small interval around the remote turning point. A substantial
reduction of computer time was achieved in the evaluations of the
quasiclassical wave functions (\ref{Uniform Langer-WKB}) and (\ref{U Fock}),
when we replaced the action integral $S_{n,j,l}^{\left(\mathrm{L}\right)}\left(r\right)$
by a corresponding (high-order) barycentric interpolation polynomials
$s_{n,j,l}^{\left(\mathrm{L}\right)}\left(r\right)$ in the interval
$0\leqslant r\leqslant r_{\mathrm{max}}$. Upon patching the wave
function of Langer and Fock inside the classically accessible region
and making use of an exact result for the normalization integral of
the Langer wave function, due to Bender and Orszag \cite{Bender-Orszag},
we finally derived an analytical result determining the quantum defect
for $l=0$ and also the value of the radial\emph{ s}-wave eigenfunctions
at the origin, thus providing a very simple and short proof of the
Fermi-Segr{\`e} formula. Also, within the range of validity of the
Fermi-contact model an analytic scaling relation for the constant
$A_{n,j,0}^{\left(\mathrm{HFS}\right)}$ describing the size of the
hyperfine shifts and splittings of the Rydberg \emph{s}-states of
the valence electron in alkali atoms was found, cf. (\ref{eq:hyperfine scaling relation}),
that apparently is for all principal quantum numbers $n\gg1$ in good
agreement with precise spectroscopic data of $^{85}\mathrm{Rb}$ and
$^{87}\mathrm{Rb}$.

\subsection*{ACKNOWLEDGMENTS}

We thank J{\'o}zsef Fort{\'a}gh, Florian Karlewski, Markus Mack,
and Jens Grimmel for useful discussions.


\begin{thebibliography}{10}
\bibitem{Marinescu et al}M. Marinescu, H. R. Sadeghpour, and A. Dalgarno,
Phys. Rev. A \textbf{49}, 982 (1994).

\bibitem{Units}We use scaled variables so that length is measured
in units of the Bohr radius $a_{B}=4\pi\varepsilon_{0}\hbar^{2}/m_{e}\left|e\right|^{2}\simeq5.2918\times10^{-11}\mathrm{m}$,
and energy is measured in units of Rydberg, $\mathrm{Ry}=m_{e}\left|e\right|^{4}/8\varepsilon_{0}^{2}h^{2}\simeq13.607\mathrm{eV}$.

\bibitem{Sanayei-Schopohl}A. Sanayei, N. Schopohl, J. Grimmel, M.
Mack, F. Karlewski, and J. Fort{\'a}gh, Phys. Rev. A \textbf{91},
032509 (2015).

\bibitem{Trefethen-Book}L. N. Trefethen, \emph{Approximation Theory
and Approximation Practice} (SIAM, Philadelphia, 2013).

\bibitem{Trefethen_Matlab}L. N. Trefethen, \emph{Spectral Methods
in MATLAB} (SIAM, Philadelphia, 2000).

\bibitem{Boyd_Book}J. P. Boyd, \emph{Chebyshev and Fourier Spectral
Methods} (Dover, New York, 2001).

\bibitem{Langer}R. E. Langer, Bull. Am. Math. Soc. \textbf{40}, 545
(1934).

\bibitem{Bender-Orszag}C. M. Bender and S. A. Orszag, \emph{Advanced
Mathematical Methods for Scientists and Engineers} (McGraw-Hill, Singapore,
1978).

\bibitem{Fock}V. A. Fock, \emph{Selected Works} (Chapman \& Hall/CRC,
New York, 2004), pp. 325-329.

\bibitem{Fermi and Segre}E. Fermi and E. Segr{\`e}, Z. Physik \textbf{82},
729 (1933).

\bibitem{supplimentary}A. Sanayei and N. Schopohl, see supplementary
material.

\bibitem{sampling theorem}W. H. Press, S. A. Teukolsky, W. T. Vetterling,
and B. P. Flannery, \emph{Numerical Recipes in C: The Art of Scientific
Computing}, 2nd ed. (Cambridge University Press, Cambridge, 1992).

\bibitem{Li et al}W. Li, I. Mourachko, M. W. Noel, and T. F. Gallagher,
Phys. Rev. A \textbf{67}, 052502 (2003).

\bibitem{Mack et al}M. Mack, F. Karlewski, H. Hattermann, S. H{\"o}ck,
F. Jessen, D. Cano, and J. Fort{\'a}gh, Phys. Rev. A\emph{ }\textbf{83},
052515 (2011).

\bibitem{Afroushed_g_states}K. Afrousheh, P. Bohlouli-Zanjani, J.
A. Pterus, and J. D. D. Martin, Phys. Rev. A \textbf{74}, 062712 (2006).

\bibitem{Gallagher_f-states}J. Han, Y. Jamil, D. V. L. Norum, P.
J. Tanner, and T. F. Gallagher, Phy. Rev. A \textbf{74}, 054502 (2006).

\bibitem{Gallagher_Book}T. F. Gallagher, \emph{Rydberg Atoms} (Cambridge
University Press, Cambridge, 1994).

\bibitem{Johnson et al}L. A. M. Johnson, H. O. Majeed, B. Sanguinetti,
Th. Becker, and B. T. H. Varcoe, N. J. Phys. \textbf{12}, 063028 (2010).

\bibitem{Blundell}S. A. Blundell, Phys. Rev. A \textbf{90}, 042514
(2014).

\bibitem{Brandenberger and Regal}J. R. Brandenberger, C. A. Regal,
R. O. Jung, and M. C. Yakes, Phys. Rev. A \textbf{65}, 042510 (2002).

\bibitem{Barut}A. O. Barut and A. Baiquni, Phys. Rev. \textbf{184},
1342 (1969).

\bibitem{Brandernberger and Malyshev}J. R. Brandenberger and G. S.
Malyshev, Phys. Rev. A \textbf{81}, 032515 (2010).

\bibitem{Bethe and Salpeter}H. A. Bethe and E. E. Salpeter, \emph{Quantum
Mechanics of One- and Two-Electron Atoms} (Springer, Berlin, 1957).

\bibitem{Froese_Fischer}C. Froese-Fischer, T. Brage, and P. J{\"o}nsson,
\emph{Computational Atomic Structure}: \emph{An MCHF Approach} (IOP
Physics, Bristol, 1997).

\bibitem{Langer_map}R. E. Langer, Phys. Rev. \textbf{51}, 669 (1937).

\bibitem{plot of higher excited states_footnote}For a plot of the
eigenfunctions $u_{n,j,l=0}(r)$ vs. $\sqrt{r}$ choosing, for example,
principal quantum numbers $n=5,9,19,29,34,39$, we refer to Ref. \cite{supplimentary}.

\bibitem{NIST}F. W. J. Olver, D. W. Lozier, R. F. Boisvert, and C.
W. Clark (eds.), \emph{NIST Handbook of Mathematical Functions} (Cambridge
University Press, New York, 2010).

\bibitem{Migdal}A. B. Migdal, \emph{Qualitative Methods in Quantum
Theory} (Addison-Wesley, Reading, MA, 1977).

\bibitem{Arimondo}E. Arimondo, M. Inguscio, and P. Violino, Rev.
Mod. Phys. \textbf{49},31 (1977).

\bibitem{Rb data sheet}D. A. Steck, \emph{Alkali D Line Data}, http://steck.us/alkalidata/
(2010).

\bibitem{Prestage et al} J. D. Prestage, R. L. Tjoelker, and L. Maleki
, Phys. Rev. Lett. \textbf{74}, 3511 (1995).

\bibitem{Durand}B. Durand and L. Durand, Phys. Rev. A \textbf{33},
2899 (1986).

\bibitem{Tauschinsky et al}A. Tauschinsky, R. Newell, H. B. van Linden
van den Heuvell, and R. J. C. Spreeuw, Phys. Rev. A \textbf{87}, 042522
(2013).\end{thebibliography}
\end{document}